\documentclass[preprint,pra,aps,showpacs,preprintnumbers,amsmath
,amssymb
]{revtex4}

\usepackage{graphicx}
\usepackage{dcolumn}
\usepackage{bm}
\usepackage{subfigure} 
\usepackage{array}
\usepackage{slashbox}
\usepackage{psfrag}
\usepackage{epstopdf}

\begin{document}

\title{Vortices in rotating optical lattices: commensurability, hysteresis, and proximity to the Mott State}

\author{Daniel S. Goldbaum}
\email{dsg28@cornell.edu}
\author{Erich J. Mueller}\affiliation{
Laboratory of Atomic and Solid State Physics, Cornell University\\
Ithaca, NY 14853
}

\date{\today}

\maketitle
\newcommand{\adi}{\hat{a}_{i}^{\dagger} }
\newcommand{\ai}{\hat{a}_{i} }
\newcommand{\adj}{\hat{a}_{j}^{\dagger} }
\newcommand{\aj}{\hat{a}_{j} }
\newcommand{\numi}{\hat{n}_{i} }

{\bf 
Quantized vortices stunningly illustrate the
 coherent nature of a superfluid Bose condensate of alkali atoms.  Introducing
  an optical lattice depletes this coherence. 
Consequently, novel vortex physics may emerge in an experiment on
 a harmonically trapped gas in the presence of a rotating optical lattice.  The most dramatic effects would occur in proximity to the Mott state, an
interaction dominated insulator with
a fixed integer number of particles per site.
 We model such a rotating gas, showing 
 that the lattice-induced spatial profile of the superfluid density drives a
  gross rearrangement of vortices.
 For example, instead of the uniform vortex lattices commonly seen in experiments,  we find parameters for which the vortices all sit at a fixed distance from the center of the trap, forming a ring.  Similarly, they can coalesce at the center, forming a
giant vortex.
We find that the
properties
of this system are hysteretic, 
even far from the Mott state.
We explain this hysteresis
in terms of vortex pinning, commensurability between vortex density and pinning site density, and energy barriers against changing the number of vortices.
Finally, we model time-of-flight expansion, demonstrating 
the experimental observability of our predictions.
}

Atomic clouds in a
 rotating optical lattice provide a system which is at the
  intersection of several major paradigms of condensed matter physics.  These rotating clouds may display a superfluid phase, an interaction driven insulating phase~\cite{Greiner:2002lr}, and even analogs of the fractional quantum Hall effect~\cite{0953-8984-20-12-123202}.  Here we explore the theory of vortices in such systems.  Most importantly, we investigate how proximity to the Mott insulator phase impacts the vortex configurations.

 Considering a uniform gas of atoms of mass $m$ 
 in an optical lattice rotating with frequency $\Omega$, there will be three
macroscopic length scales in the problem: the lattice spacing $d$, the  magnetic length $\ell=\sqrt{\hbar/m\Omega}$, and the particle spacing $n^{-1/3}$, where $\hbar=h/2\pi$ is Planck's constant.  Even without interactions, the commensurability of these various lengths can lead to nontrivial physics -- the single particle spectrum, the Hofstadter butterfly, is fractal \cite{PhysRevB.14.2239}.  For a system of interacting bosons, this fractal spectrum leads to a modulation of the boundary between  superfluid and Mott insulating phases~\cite{goldbaum:033629,umucalilar:055601}.  Further, the vortices in a superfluid on a rotating lattice develop extra structure: for example they can have cores filled with the Mott state~\cite{wu:043609}.  Related to this structure, the vortices can become pinned at optical lattice potential minima, in contrast to the more usual pinning at potential maxima~\cite{goldbaum:033629}.  We explore how this physics plays out in a finite system, as can be studied in experiments.

We consider a harmonically trapped gas of bosons in a rotating two-dimensional square lattice.  
A recent experiment~\cite{tung:240402} realized exactly this scenario by placing a rotating mask in the Fourier plane of a laser beam which forms an optical dipole trap.  The mask contained three/four holes, producing a triangular/square lattice in the image plane, where the atoms were trapped.  The lattice spacing was large due to the nature of their optics but
can 
be made comparable to the wavelength of the trapping light.
If one reduced this lengthscale, one could explore the tight binding limit, where atomic motion is limited to the lowest band, and Mott insulating physics could become important.  We consider the theory of this tight binding limit.  This limit may also be reached through
 quantum optics techniques which introduce phases on the hopping matrix elements for atoms in a non-rotating lattice~\cite{mueller:041603, sorensen:086803, JakschNJP2003, palmer-2008}.  We choose to study a two-dimensional cloud, as it provides the simplest setting for investigating vortex physics. 
 This is also an experimentally relevant geometry, as
 the dimensionality of the system can be  controlled by using an anisotropic harmonic potential, or optical lattice, where the hard trapping direction is along the optical lattice rotation axis~\cite{spielman:120402}.

In the tight-binding limit, this system is described by 
the rotating Bose-Hubbard hamiltonian~\cite{wu:043609,PhysRevLett.81.3108}:
\begin{equation}
\hat{H}
=-\sum_{\langle i,j \rangle} \left(t_{ij} \adi  \aj  + h.c. \right) 
+  \sum_i \left(\frac{U}{2} \numi \left( \numi - 1 \right) -\mu_i \numi\right)
\label{B}
\end{equation}
where $t_{ij}=t\exp\left[ i \int_{\vec{r}_j}^{\vec{r}_{i}} d\vec{r} \cdot \vec{A}(\vec{r}) \right]$ is the hopping matrix element from site $j$ to site $i$.  The rotation vector potential, which gives rise to the Coriolis effect, is $\vec{A}(\vec{r}) = \left( m/\hbar\right) \left(\vec{\Omega} \times \vec{r} \right)= \pi \nu \left( x \hat{y} -y \hat{x} \right)$, where $\nu$ is the number of circulation quanta per optical-lattice site.  The local chemical potential
 $\mu_i=\mu_0-m \left( \omega^2-\Omega^2 \right) r_i^2 /2$ includes the centripetal potential.  In these expressions,
 $\mu_0$ is the central chemical potential, $\omega$ is the trapping frequency, $\Omega$ is the rotation speed, ${\vec r}_i$ is the position of site $i$,  $m$ is the atomic mass, $\adi$ $\left( \ai \right)$ is a bosonic creation (annihilation) operator,
$\numi = \adi \ai $ is the particle number operator for site $i$,  and $U$ is the particle-particle interaction strength.      
The connection between these parameters and the laser intensities are given by Jaksch et al. \cite{PhysRevLett.81.3108}.
Here, and in the rest of the paper, we use units where the lattice spacing is unity.

As described in the Methods section, we use the variational Gutzwiller \emph{ansatz} to calculate the properties of this system.  This approach describes well both the superfluid and Mott insulating states, and has the important property that the superfluid density $\rho^c$ is not equal to the total density $\rho$.  This is one of the key ways in which the lattice system differs from a traditional weakly interacting gas of bosons, and much of the physics we see is contingent upon this feature.

We systematically explore the phase space, varying the parameters in the hamiltonian.  
We simulate clouds with diameter from 30-60 sites, comparable to the sizes studied in experiments~\cite{GretchenK.Campbell08042006, folling:060403}.  For the largest simulations we impose four-fold rotational symmetry, but introduce no constraints in the smaller simulations.
We find interesting results both near and far from the Mott regime.  For weak interactions the dominant physics involves the pinning of vortices between lattice sites, while near the Mott boundary, the non-uniform superfluid density becomes central.

\section*{Commensurability, Hysteresis, and Pinning}

 Previous work, focusing on the multi-band weak lattice limit, found results similar to those we see in our tight binding model far from the Mott regime \cite{pu:190401,reijnders:060401}.  We find a great variety of vortex structures, including patterns resembling square vortex lattices.  These are most stable at the rotation rates where they are commensurate with the underlying optical lattice. 
Commensurate Bravais lattices exist when $1/\nu$ is an integer, and commensurate square lattices when $\nu=1/(n^2+m^2)$, for integral $n$ and $m$ \cite{tung:240402,reijnders:060401,pu:190401}.  Which vortex patterns appear in a simulation, or in an experiment~\cite{PhysRevLett.86.4443}, depends on how the system is prepared.
This hysteresis occurs because
 the energy landscape 
 has many deep gorges with near-degenerate energies, separated by high barriers.
 
\begin{figure}
\includegraphics[width=0.45\columnwidth]{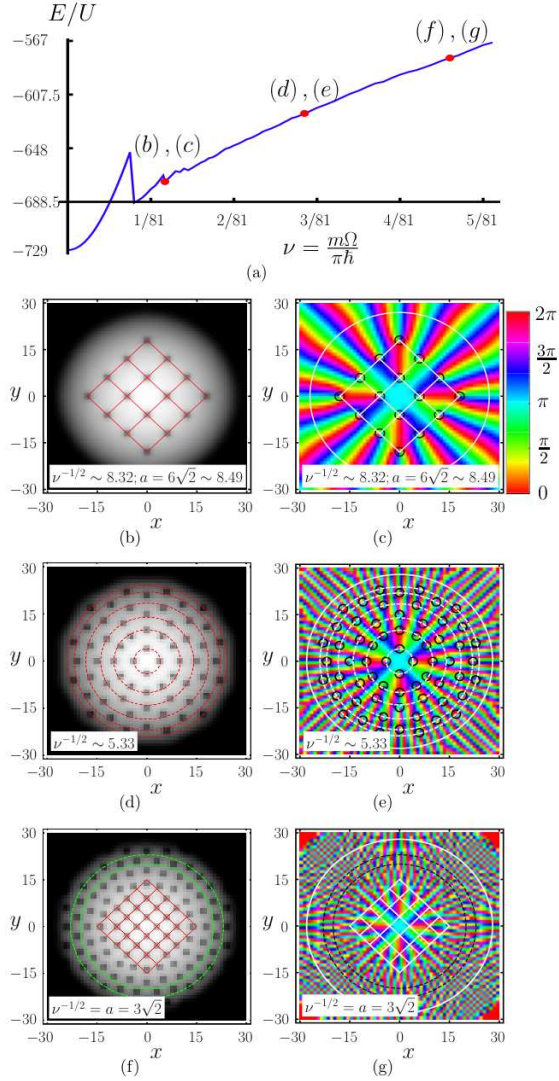}
\caption{\textbf{Adiabatic spin-up} (color, one-column).  
Properties of cloud during adiabatic spinup with parameters
 $\left(t/U=0.2, \, \mu_0/U=0.3, \, R_{TF}=15 \right)$. (a) Energy vs rotation rate. Sharp drops indicate vortex formation. Energy scaled by on-site interaction parameter $U$, and rotation rate quoted as $\nu$, the number of circulation quanta going around each plaquette. (b), (d), (f) Superfluid density profile at parameters labeled in (a).  Dark-to-light shading corresponds to low-to-high density, and position is measured in lattice spacing. Dark spots correspond to vortex cores. Red and green lines are guides to the eye. (c), (e), (g)  Superfluid phase represented by Hue.  Solid white circle denotes edge of cloud.  Dashed lines are guides to the eye.  Black circles denote vortex locations.  In (b), (c) and (f), (g) rotation speed should favor a commensurate square vortex lattice rotated by $\pi/4$ from the optical lattice directions. (d), (e) represents an incommensurate rotation speed.
 }
\label{V-Lattice_Figure_Prep}
\end{figure}

 To illustrate this energy landscape, 
 we fix $t/U=0.2$ and $\mu_0/U=0.3$, and study how the energy evolves as we change the rotation speed.
First, starting with the non-rotating ground state, we sequentially increase the rotation speed, using the previous wavefunction as a seed for our iterative algorithm.  We adjust $\omega$ as we increase $\Omega$ so that the cloud size, related to the Thomas-Fermi radius, $R_{TF}=\sqrt{\frac{2 \mu_0}{m\left( \omega^2-\Omega^2 \right)}}$, remains effectively fixed.  
The energy as a function of rotation speed, shown in Figure~\ref{V-Lattice_Figure_Prep}(a), has a series of sharp drops, corresponding to the entry of one or more vortices from outside of the cloud.  At these rotation speeds the system jumps from one local minimum of the energy landscape to another. 

 Figure \ref{V-Lattice_Figure_Prep} (b)-(g) shows the superfluid density and phases associated with the vortex patterns found during this {\em adiabatic} increase in rotation speed, where we impose four-fold rotation symmetry about the trap center. 
Subfigures
 (b) and (c) show a regular square vortex lattice 
 seen near the commensurate $\nu=1/(2\times6^2)$.  Subfigures (d) and (e) show the vortex configuration at $\nu\sim1/(2\times 3.76^2)$ which is intermediate between the commensurate values
$\nu=1/(2\times 3^2)$ and  $\nu=1/(2\times 4^2)$.  Rather than forming a square pattern, the vortex configuration appears to be made of concentric rings.  Such ring-like structures also occur for superfluids rotating in hard-walled cylindrical containers \cite{PhysRevB.20.1886}, where boundaries play an important role.  Changing the shape of the boundary should change the vortex configuration.  For example, by using an asymmetric trap one should produce elliptical shells of vortices.  As one increases $\nu$ towards  $\nu\sim 1/(2\times 3^2)$, a domain containing a square vortex lattice begins to grow in the center of the trap.  As  seen in subfigures (f) and (g), at commensurability the 
 domain only occupies part of the cloud, even though one would expect that a uniform square lattice would be energetically favorable.  The inability of the system to find the expected lowest energy configuration during an adiabatic spin-up is indicative of the complicated energy landscape.

The patterns which we find are largely determined by the symmetry of the instabilities by which vortices enter the system.  For example, even when we do not impose a four-fold symmetry constraint this adiabatic spin-up approach never produces square vortex lattices oriented at an angle other than $\pi/4$ with respect to the optical lattice axes.  On the other hand, we readily produce other commensurate vortex lattices by choosing the appropriate rotation speed and seeding our iterative algorithm with the corresponding phase pattern.      We have verified this approach with square vortex lattices oriented at various angles with respect to the optical lattice, taking $\nu^{-1/2}=\{4,5,6,7,8,9,10\}$,  $(5 \nu)^{-1/2}=\{2,3,4,5,6\}$ and $(10 \nu)^{-1/2}=\{2,3,4\}$.

\begin{figure}
\includegraphics[width=0.45\columnwidth]{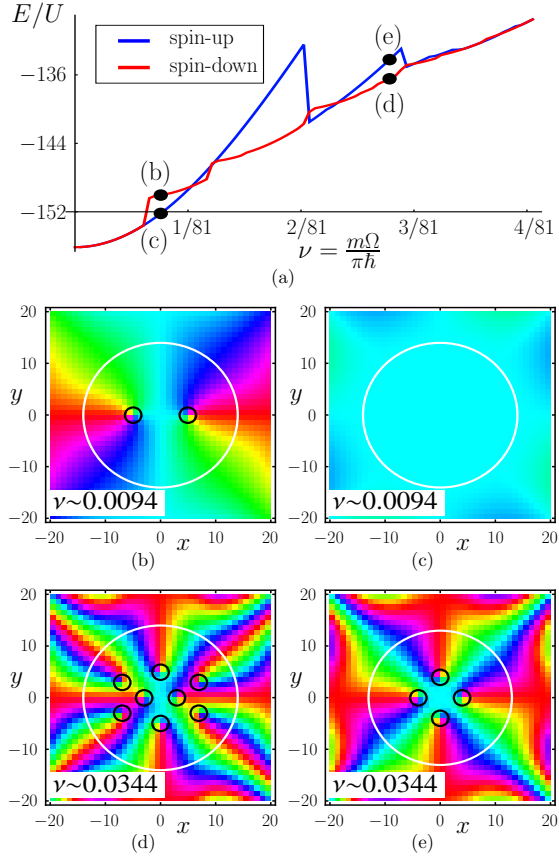}
\caption{\textbf{Hysteresis} (color, one-column). (a) Energy  versus rotation rate during increase (blue line) and decrease (red line) of $\nu$.  Energy steps in the blue (red) curve correspond to nucleation (expulsion) of vortices. (b)-(e) Order parameter complex phase for parameters labeled in (a). Black circles are drawn around vortex cores, and white circles indicate the approximate extent of the gas. 
}
\label{Hysteresis}
\end{figure}

We further explore the history dependance of the vortex configurations
by  increasing, then decreasing $\Omega$.
We
do not impose a four-fold symmetry constraint, but take a smaller system with $R_{TF}=7$.
    At any given $\Omega$, the energy shown in figure~\ref{Hysteresis} (a) depends on the system's history.  The increasing(blue)/decreasing(red) rotation curve has sharp energy drops signaling the introduction/ejection of vortices to/from the system.  The energy drops occur at different $\Omega$ for spin-up and spin-down, indicating that the critical rotation speed for a vortex to enter or exit the system is different.   Generically, there are more vortices in the system on spin-down than on spin-up.  Depending on $\Omega$, one may find a lower energy state by increasing  (subfigs. (b) and (c)) or by decreasing (subfigs. (d) and (e)) the rotation rate.
As demonstrated by subfigs. (d) and (e), vortex configurations produced during spin-up typically have the four-fold rotational symmetry of the optical lattice, while the vortex configurations calculated during spin-down are more likely to break this symmetry.  An experiment will display the same qualitative features, but slightly different vortex configurations. 

\section*{Influence of Mott physics}

\begin{figure}
\includegraphics[width=0.45\columnwidth]{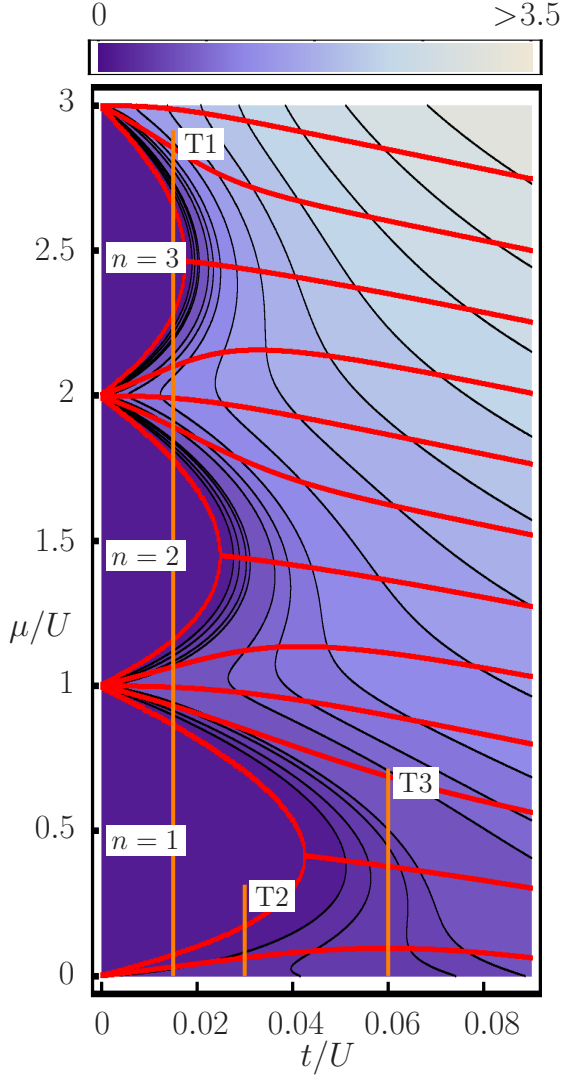}
\caption{\textbf{Mean-field 2D Bose-Hubbard model phase diagram} (color, one-column). 
Red (black) curves:  contours of constant total-particle (superfluid) density. Dark-to-light purple shading: low-to-high superfluid density. Qualitative understanding of trapped gas is produced by following lines of fixed $t/U$, such as those illustrated by orange lines  \{T1, T2, T3\} discussed in the text. 
}
\label{UniformContours}
\end{figure}

Next we investigate the system in a deeper optical lattice, close enough to the Mott insulator phases so that 
$\rho^c$ and $\rho$ are noticeably different.
A basic understanding of this region comes from investigating the phase diagram of the homogeneous system, where $\Omega=\omega=0$.  The mean-field phase diagram for the uniform non-rotating system is shown in Fig.~\ref{UniformContours}.  Contours of fixed $\rho$ and $\rho^c$, are indicated by red and black curves. The superfluid density vanishes in the Mott regions where the density is commensurate with the lattice, and the number of particles per site labels each lobe.
 For a sufficiently gentle trap, the gas looks locally homogeneous, and its density at any point $r$ can be approximated by that of a uniform system with chemical potential $\mu(r)=\mu_0-V(r)$.  We go beyond this local density approximation (LDA) in our calculations, however it is useful for gaining qualitative insight into the density profiles we expect.
 
For illustration, consider the 
 three vertical (fixed $t/U$) lines in figure~\ref{UniformContours}, corresponding to the LDA trajectories of the chemical potential in three separate trapped clouds. Trajectory ``T1'' represents a system consisting of alternating superfluid and Mott-insulator phases. Trajectory ``T2'' represents a Mott-insulating core surrounded by a superfluid ring. Trajectory ``T3'' represents a system with a high $\rho^c$ central region surrounded by a plateau of near constant $\rho^c$.  We self-consistently calculate the density profile of atomic clouds where $\mu_0/U$ corresponds to the tops of the T3 and T2 lines.

\begin{figure}
\includegraphics[width=0.45\columnwidth]{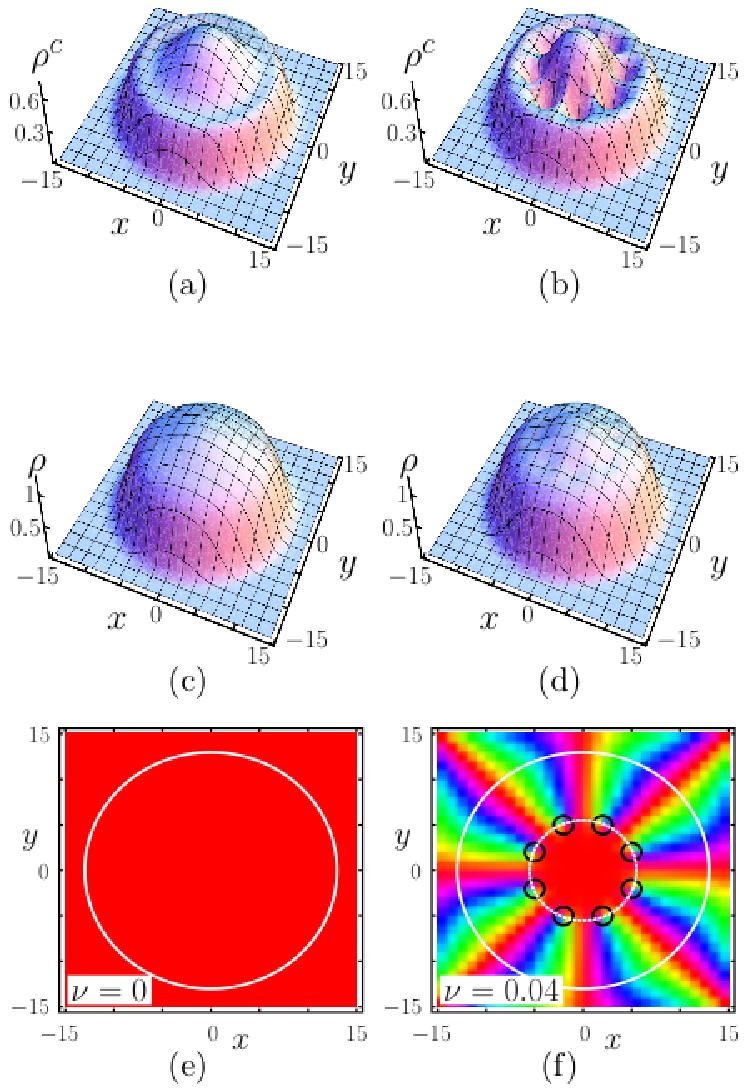}
\caption{\textbf{Ring vortex configuration} (color, one-column). (a), (c), (e) Superfluid density, total-particle density, and complex phase profile for a non-rotating system at $\left(t/U=0.06,\,  \mu_0/U=0.7 ,\, R_{TF}=10\right)$ corresponding to T3 in figure~\ref{UniformContours}. (b), (d), (f) Same after adiabatically increasing rotation rate to $\nu=0.04$. The vortices sit in a circle surrounding the central high superfluid density region. The central fluid is approximately stationary with constant phase, while the outer velocity is approximately azimuthal. }
\label{RingFigurePrep}
\end{figure}

In figure~\ref{RingFigurePrep} we display the density and complex phase profiles corresponding to T3 for the non-rotating and rotating optical lattice cases. These calculations are performed at $\left(t/U=0.06, \mu_0/U=0.7, R_{TF}=10
\right)$, and with four-fold rotation symmetry enforced. The cloud has a diameter of roughly 26 sites.  In the non-rotating case we see that the  superfluid density profile (subfigure (a)) has the plateau structure predicted by the LDA. No such plateau is seen in the total density (subfigure (c)).

 The phase of the superfluid order parameter is uniform in the absence of rotation (subfigure (e)).
Starting from this non-rotating configuration we gradually increase the rotation speed to 
  $\nu=0.04$.  Rather than forming a square lattice, the resulting vortices form a ring around the central $\rho^c$ peak (subfigure (c)).   
  This  configuration is  favored because it minimizes the sum of competing energy costs:
  the rotation favors a uniform distribution of vortices, but the single vortex energy is smallest where $\rho^c$ is low.

As seen in subfigure (f), the phase of  the superfluid inside the ring is essentially constant.  This can be understood by an analogy with magnetostatics.
  Given that the velocity field around a vortex line, $\vec{\bold v}$, is analogous to the magnetic field around a current carrying wire, it obeys the equivalent of Ampere's law $\oint \vec{\bf  v}\cdot d\vec{\bf \ell}=(h/m) N_v$, where $N_v$ is the number of vortices enclosed in the contour of integration.  In the limit where one can neglect the discreteness of the vortices in the ring, one finds that the fluid inside is motionless, while the fluid outside moves with the same azimuthal velocity that one would find if all the vortices were at the geometric center of the cloud.
Even with only eight vortices our system appears to approach this limit.      
If one increased the rotation speed, one
could imagine finding a state with several concentric rings of vortices in the plateau.  
Similarly, increasing $\mu_0/U$ could produce multiple superfluid plateaus, each of which could contain a ring of vortices.  This structure is reminiscent of Onsager and London's original proposal of vortex sheets in liquid helium \cite{london}.

\begin{figure}
\includegraphics[width=0.45\columnwidth]{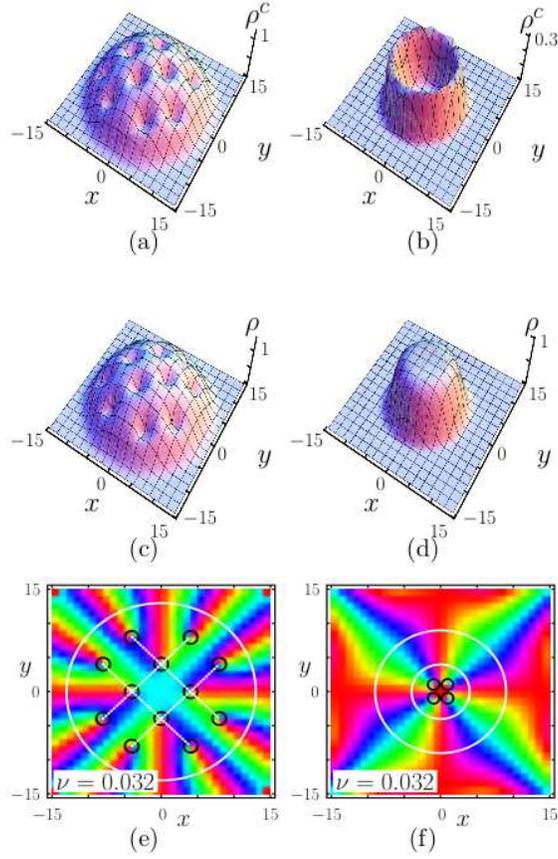}
\caption{\textbf{Giant vortex} (color, one-column). (a), (c), (e)  Superfluid density, total-particle density and complex phase field for a square vortex-lattice state prepared by adiabatically increasing rotation rate at relatively low optical lattice depth $\left(t/U=0.2, \mu_0/U=0.3,\,  R_{TF}=7,\, \nu= 0.032\right)$.  (b), (d), (f) Same after adiabatically increasing optical lattice depth so that $t/U=0.03$.  A superfluid ring surrounds a Mott state, and four of the vortices have coalesced to form a giant vortex of charge 4.}
\label{GiantFigurePrep}
\end{figure}

In figure~\ref{GiantFigurePrep} we investigate the state corresponding to trajectory T2, where the LDA predicts a superfluid shell surrounding a Mott core.   
Rather than forming a lattice of discrete vortices, one expects that this system will form a ``giant" vortex when rotated: the vortices occupying the Mott region, leaving a persistent current in the superfluid shell.  The energy barriers for changing vorticity should be particularly high, so we generate the rotating state in two stages.  (A similar protocol could be used in an experiment.)  We start with a non-rotating system ($\nu=0$) at weak coupling $\left( t/U=0.2, \mu_0/U=0.3 \right)$, gradually increasing the rotation to $\nu=0.032$, where we find the square vortex lattice illustrated in subfigures (a), (c) and (e).  We then adiabatically reduce  $t/U$ from $0.2$ to $0.03$.  As we reduce $t/U$, the central $\rho^c$ drops, while $\rho$ approaches unity there.  Eventually we see a Mott regime at the center of the cloud.  During the evolution, we find that 8 of the vortices migrate to the edge of the trap and then escape, while four of the vortices move towards the center.
These four vortices coalesce at the center of the trap and effectively form a vortex of charge 4, which is illustrated in subfigures (b), (d) and (f).  Such a dense packing of vorticity would be unstable in the absence of the optical lattice.  For larger systems with higher rotation rates we have produced giant vortices with much higher quantization.

Due to its multiply connected topology, a ring is one of the archetypical geometries used in theoretical discussions of superfluidity~\cite{1991LNP...394....1L,rokhsar-1997,PhysRevA.57.R1505}.  There are several experimental schemes for creating a ring-shaped trap (for example using the dipole forces from a blue detuned laser focussed at the center of an atomic cloud) \cite{ryu:260401}, and many theoretical studies of giant vortex formation stabilized by a quadratic-plus-quartic potential~\cite{PhysRevA.66.053606,jackson-2004-69,aftalion-2004-69,jackson-2004-70,bargi-2006-73}.  Here the multiply connected geometry is spontaneously formed by the appearance of the Mott state in the center of the cloud. As was found in a related study by Scarola and Das Sarma~\cite{scarola:210403}, this Mott region effectively pins the vortices to the center.  

By changing $t/U$ one may study a few other interesting structures.  For example, one can engineer a situation where a central superfluid region is surrounded by a Mott ring followed by a superfluid ring.  At appropriate rotation speeds one produces a configuration which has properties of both the states seen in figure~\ref{RingFigurePrep} and in figure~\ref{GiantFigurePrep}.   One will find no vortex cores (all of the vorticity is confined to the Mott ring), the central region will be stationary, and  the outer region rotates.  

Another interesting limit is found when one decreases the  thickness of a Mott/superfluid region so much that it break up into a number of discrete islands.  Small Mott islands act as pinning centers, while small superfluid islands form an analog of a Josephson junction array~\cite{Fazio:2001ta, Cataliotti:2001wc}.

\section*{Detection}

Many of the structures we have discussed would be very hard to detect using {\em in-situ} absorption imaging.
As is exemplified by figure 4(d), the vortices do not necessarily have a great influence on the density of the cloud.  This is principally because near the Mott boundary the superfluid fraction becomes small:  even though the superfluid vanishes in the vortex core, this does not appreciably change the density.   
Two other pieces of physics also influence the visibility.  First, near the Mott boundary one can produce vortices with Mott cores~\cite{wu:043609}.  Depending on the bulk density, this can lead to vortices where there is no density suppression at all, or even a density enhancement.  Second, the lengthscale of the vortex core, the superfluid healing length, varies with $U/t$.  For both very large and very small $U/t$ the healing length is very large, while at intermediate couplings it is comparable to the lattice spacing, possibly below the resolution of typical optics. 

\begin{figure}
\includegraphics[width=0.45\columnwidth]{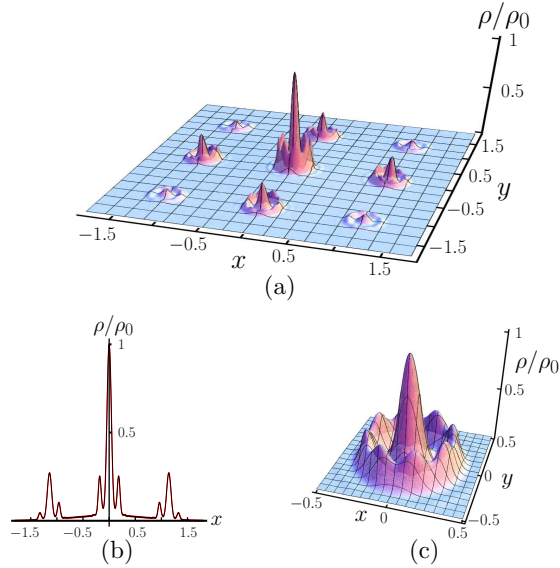}
\caption{\textbf{Time-of-flight expansion} (color, one-column). (a) Column density $\rho$ scaled by the central column density $\rho_0$ as a function of space for the state in fig.~\ref{RingFigurePrep} (b) , (d), (f) after expanding for time $t$.  Positions are measured in terms of scaling parameter $D_t=\hbar t/m \lambda$, where $\lambda$ is the size of the initial Wannier state in each site of the lattice.  This scaling solution holds in the long time limit where $D_t$ is large compared to the initial size of the cloud. (b) One dimensional cut through center of (a).  Note the incoherent background between the Bragg peaks. (c) Close-up of the central Bragg peak, corresponding to the Fourier transform of the superfluid order parameter.
There are 8 dips in the outer crest, corresponding to the 8 vortices in the initial state.
}
\label{TOFFig}
\end{figure}

We argue that the vortex structures will be much more easily imaged after time-of-flight expansion of the cloud~\cite{PhysRevLett.84.806, ketterle-1999}.  The density after time-of-flight expansion is made of two pieces.  There will be a largely featureless incoherent background from the normal component of the gas, and a coherent contribution from the superfluid component.  Due to interference from the different lattice sites, the coherent contribution will form a series of Bragg peaks~\cite{Greiner:2002lr}.  Each peak will reflect the Fourier transform of the superfluid order parameter.
In the methods section we present a simple approach, neglecting interactions during the time-of-flight, which allows us to calculate the time-of-flight expansion images.  Figure~\ref{TOFFig} illustrates the density pattern which will be seen if the rotating cloud in fig.~\ref{RingFigurePrep} is allowed to expand. 

\section*{Summary}

In this paper we have studied the vortex configurations in a harmonically trapped Bose gas in the presence of a rotating deep optical lattice. Far from the Mott phases we illustrated the effects of vortex-pinning and commensurability in determining ground state vortex configurations. We studied the hysteretic evolution of vortex configurations as the rotation speed is varied.  We explored the influence of the Mott phase on the rotating superfluid, encountering a series of novel vortex structures such as rings and giant vortices.  Finally, we explained how these structures can be observed in time-of-flight images of the atomic cloud.

\section*{Methods}
\subsection*{General}
Both the superfluid and Mott insulator are well described by 
a spatially inhomogeneous Gutzwiller product \emph{ansatz}~\cite{PhysRevLett.81.3108},
\begin{equation}
\lvert \Psi_{GW} \rangle = \prod_{i=1}^{M} \left( \sum_n f_n^i \lvert n \rangle_i \right) \, ,
\label{C}
\end{equation}  
where $i$ is the site index, $M$ is the total number of sites, $\lvert n \rangle_i$ is the n-particle occupation-number state at site $i$, and $f_n^i$ is the corresponding complex amplitude, with $\sum_n |f_n^i|^2=1$.
This ansatz does not capture the correlations found in analogs of fractional quantum Hall states, and hence we do not explore this model in that regime (which is characterized by the interparticle spacing being of order the magnetic length).  
As a mean-field approach, it also does not capture quantum fluctuation effects seen near the critical point \cite{PhysRevB.40.546, capogrosso-sansone:134302}, nor is it appropriate for describing strongly-correlated macroscopic superposition states~\cite{rey-2008}.  
Furthermore it neglects all short-range correlations.
Despite these limitations, the Gutzwiller approach compares well with exact methods, and strong coupling expansions \cite{ PhysRevB.40.546, capogrosso-sansone:134302, PhysRevB.53.2691}.  It
has been used extensively to understand experimental results~\cite{folling:060403,GretchenK.Campbell08042006, Greiner:2002lr}, and is well suited for studying the vortex physics that we consider here.

Using equation~\eqref{C} as a variational {\em ansatz}, we minimize the energy with respect to the $\{f_n^i \}$.  We then extract the density $\rho_i=\sum_n n |f_n^i|^2$ and the condensate order parameter $\alpha_i=\langle \hat{a}_i \rangle= \sum_n \sqrt{n} \left(f_{n-1}^{i}\right)^* f_n^i$ at each site. The condensate density $\rho_i^c=\lvert \langle \hat{a}_i \rangle \rvert^2$ is equal to the superfluid density in this model, and is generally not equal to the density.

We use an iterative algorithm to  determine the $\{f_n^i\}$ which minimize the energy.  We use a square region with $L$ sites per side with hard boundary conditions. We find that we must take 
$L$ much larger than the effective trap radius so that our solutions do not depend on those boundaries.  Typically we use $40\leq L \leq 90$.
We calculate $\langle \hat{H}_{RBH} \rangle$ using equation \eqref{C}.  Minimizing $\langle \hat{H}_{RBH} \rangle$ with respect to $f_n^{i*}$ with the constraint $\sum_n |f_n^i|^2-1=0$ gives $L^2$ nonlinear eigenvalue equations, one for each site,
\begin{equation}
 -t \sum_{k \text{, nn of } j}  \left( \langle \hat{a}_k \rangle  \sqrt{m}  f_{m-1}^j R_{jk} + \langle \hat{a}_k^{\dagger} \rangle \sqrt{m+1} f_{m+1}^j R_{kj} \right) 
+ \left( \frac{U}{2} m^2- \left( \mu\left( r \right)+\frac{U}{2} \right) m + \lambda_j \right) f_m^j = 0 \, ,
\label{M}   
\end{equation}
where the sum is over all nearest neighbors of site $j$, $m$ is the particle-number index, $\lambda_j$ is a Lagrange multiplier, and $R_{jk} = \exp{ \left[ i \int_{\bold{r}_k}^{\bold{r}_{j}} d\bold{r} \cdot \bold{A}(\bold{r}) \right]}$, where $i=\sqrt{-1}$. We iteratively solve these equations: first choosing a trial order-parameter field $\left\{ \alpha_j^{\left( 0 \right)} \right\}$, where $ \alpha_j = \langle \hat{a}_j \rangle $; then updating it by $\alpha_j^{\left( p \right)} = \sum_n \sqrt{n} f_{n-1}^{j*} \left( \left\{ \alpha_j^{\left( p-1 \right)} \right\} \right) f_n^j \left( \left\{ \alpha_j^{\left( p-1 \right)} \right\} \right)$, where $p$ is the iteration index. Similar calculations were performed by Scarola and Das Sarma~\cite{scarola:210403} to analyze the case where the single-particle Mott state is surrounded by a rotating superfluid ring.

Since equation (\ref{M}) is highly nonlinear, we find that the solution that this iterative algorithm converges to is sensitive to the initial state we use.  This feature allows us to see the hysteretic effects described in the text.  Experiments will see similar hysteresis, but the precise details will differ from our calculations (for example the critical frequencies for vortex entry and egress will be somewhat modified).

\subsection*{Detection}
To detect the vortices we propose time-of-flight imaging~\cite{Greiner:2002lr,PhysRevLett.84.806}, where at time $t=0$ one turns off the lattice and the harmonic trap, letting the cloud expand.  After some fixed time $t$ one then produces an absorption image of the cloud using a resonant laser beam.  In a weakly interacting gas, the density profile is related to the momentum distribution of the gas.  As we argue below, the vortices will be observable in the time-of-flight images.

To illustrate, we present a simple model where we neglect interactions during the time of flight.  As applied to atoms expanding from different sites, this approximation should be sound:  by the time atoms from different sites overlap, the density is so low that they have negligable chance of scattering.  On the other hand, atoms from the same site do scatter off one-another.  These interactions have two effects: (1) atoms from sites with higher occupation will be moving faster (the interaction energy is converted into kinetic energy), and (2) the interactions introduce phase shifts which depend on atom number.  Both of these effects will slightly degrade the contrast of the resulting interference pattern.  When the interaction energies are small compared to the band spacing (a condition required for the single band Hubbard model to be applicable) these effects will be small.

Within our approximation, calculating the expansion reduces to a series of single-particle problems.  Taking the initial wavefunction to be given by equation \eqref{C}, after time $t$  the wavefunction will be
\begin{equation}
|\psi(t)\rangle =\prod_{i=1}^{M}  \left(\sum_n f_n^i  \frac{\left[\hat a_i^\dagger(t)\right]^n}{\sqrt{n!}}\right)\lvert {\rm vac} \rangle,
\end{equation}
where $\hat a_i(0)$ is the operator which annihilates the single-particle state in site $i$ of the lattice.  This operator evolves via the Heisenberg equation of motion, 
$
i \hbar \partial_t \hat a_j(t)=\left[ \hat a_j(t),\hat{H}_{\rm{free}} \right]
$
where $\hat{H}_{\rm{free}}$ is the Hamiltonian for non-interacting particles.  This is equivalent to evolving the single particle state annihilated by $\hat a_i(t)$ via the free Schrodinger equation.

For this analysis we use the notation that $\vec{\bold{r}}$ is a vector in the $x-y$ plane, and $z$ represents the coordinate in the perpendicular direction.
We take the initial (Wannier) state at each site, $\phi_i \left( \vec{\bold{r}},z \right)$, to be gaussian:
\begin{equation}
\phi_i \left( \vec{\bold{r}},z \right)=\frac{1}{\left( \pi \lambda^2 \right)^{1/2}}\frac{1}{\left( \pi \lambda_\perp^2 \right)^{1/4}} \exp{\left[ -\frac{\left(\vec{\bold{r}}-\vec{\bold{r_i}}\right)^2}{2 \lambda^2}-\frac{z^2}{2\lambda_\perp^2} \right]} \, ,
\label{N}
\end{equation}
where $\lambda=\sqrt{\frac{\hbar}{m \omega_{osc}}}$, and $\lambda_\perp=\sqrt{\frac{\hbar}{m \omega_\perp}}$ with $\omega_{osc}$ and $\omega_\perp$ being the small oscillation frequencies for each well. In the geometry we envision, $\omega_\perp\gg \omega_{osc}$.  The wavefunctions at a time $t$ after release of the trap are calculated by Fourier transforming $\phi_i \left( \vec{\bold{r}},z \right)$ to momentum space, then time evolving under $\hat{H}_{\rm{free}}$ and finally Fourier transforming back to position space to arrive at $\phi_i \left( \vec{\bold{r}},z,t \right)=\phi_i \left( \vec{\bold{r}},t \right) f(z,t)$,  where the only thing we need to know about the transverse wavefunction $f(z,t)$ is that it is normalized so $\int |f(z,t)|^2 dz=1$.  The in-plane wavefunction is
\begin{equation}
\phi_i \left( \vec{\bold{r}},t \right)= \left( \frac{\lambda^2}{\pi \left( \lambda^2+i \hbar t/m\right)^2}\right)^{1/2} \exp{\left[ -\frac{\left(\vec{\bold{r}}-\vec{\bold{r_i}}\right)^2}{2 \left( \lambda^2+i \hbar t/m \right)}\right]},
\label{O}
\end{equation}
The column density of the expanding cloud is then
 \begin{equation}
 n\left( \vec{\bold{r}},t \right)=
 \int \langle \psi(t)| \hat{\psi}^{\dagger}\negmedspace \left( \vec{\bold{r}},z \right)  \hat{\psi} \negmedspace \left( \vec{\bold{r}},z \right) |\psi(t)\rangle\, dz= \sum_{i=1}^M \left[ n_i - n_{c,i} \right] \lvert \phi_i \left( \vec{\bold{r}},t \right) \rvert^2 + \Bigg| \sum_{i=1}^M  \alpha_i \phi_i \left( \vec{\bold{r}},t \right) \Bigg|^2 \, ,
 \end{equation}
 where $n_i \, \, \left( n_{c,i} \right)$ is the number of atoms (condensed atoms) initially at site $i$, and $\hat{\psi} \negmedspace \left( \vec{\bold{r}},z \right)$ is the bosonic field operator annihilating an atom at position $(\vec{\bold{r}},z)$. 
 
 In the long time limit where  the expanded cloud is much larger than the initial cloud (ie. $D_t=\hbar t/m\lambda\gg R_{TF}$), this expression further simplifies, and one has
 \begin{eqnarray}
n(r,t)&=&
\rho(r,t)
\left[ 
(N-N_c) 
+ |\Lambda(r,t)|^2
\right]
\\
\rho(r,t)&=&\left(\pi D_t^2 \right)^{-1} e^{-r^2/D_t^2}
\\
\Lambda(r,t)&=&\sum_j \alpha_j e^{-i {\bf r \cdot r}_j/D_t \lambda},
\end{eqnarray}
where 
$N$ and $N_c$ are the total number of particles and condensed particles, respectively.
The envelope, $\rho(r,t)$, is a Gaussian, reflecting the Gaussian shape of the Wannier state. 
 The incoherent contribution $(N-N_c) \rho(r,t)$
 
 has no additional structure. This is a consequence of the Gutzwiller approximation, which neglects short range correlations.
 Adding these correlations would modify the shape of the background, but it will remain smooth.

The interference term has the structure of the envelope $\rho(r,t)$ multiplied by the modulus squared of the discrete Fourier transform of the superfluid order parameter.  The discrete Fourier transform  can be constructed by taking the continuous Fourier transform of the product of a square array of delta-functions, and a smooth function which interpolates the superfluid order parameter.  The resulting convolution
produces
 of a series of Bragg peaks, each of which have an identical internal structure which is the  Fourier transform of the interpolated superfluid order parameter.  The vortices will be visible in the structure of these peaks.
 
 Vortices in real-space lead to vortices in reciprocal space.  This result is clearest for ``lowest Landau level" vortex lattices~\cite{PhysRevLett.87.060403} which are expressible as an analytic function of $z=x+i y$ multiplied by a Gaussian of the form $e^{-|z|^2/w^2}$, where $w$ is a length scale which sets the cloud size.  Aside from a scale factor and a rotation, the continuous Fourier transform of such a function is identical to the original.  More generally, the topological charge associated with the total number of vortices is conserved in the expansion process.
 
 Figure~\ref{TOFFig} shows a sample expansion image in the long expansion limit, where up to a scaling of the spatial axes, the density pattern only depends on the $\{f_n^i\}$'s from the Gutzwiller ansatz wavefunction, and the ratio between the size of the Wannier states and the lattice spacing ($\lambda/d$).  We use the initial state shown in fig.~\ref{RingFigurePrep}, where $t/U=0.06$.  Using $d=410$nm and hard axis lattice depth of $30 E_R$, which are the experimental values in ref.~\cite{spielman:120402}, we find that $V_0=9.3 E_R$, which gives $\lambda=75$nm.

\section*{Acknowledgements}
We thank Joern Kupferschmidt for 
discussions about the detection method, and for
performing some preliminary time-of-flight simulations. We thank Kaden Hazzard for providing code to calculate the Bose-Hubbard parameters from the underlying continuum model. We also thank Bryan Daniels and Kaden Hazzard for illuminating discussions.
This material is based upon work supported by the National Science Foundation through grant No. PHY-0758104.


\begin{thebibliography}{10}
\expandafter\ifx\csname url\endcsname\relax
  \def\url#1{\texttt{#1}}\fi
\expandafter\ifx\csname urlprefix\endcsname\relax\def\urlprefix{URL }\fi
\providecommand{\bibinfo}[2]{#2}
\providecommand{\eprint}[2][]{\url{#2}}

\bibitem{Greiner:2002lr}
\bibinfo{author}{Greiner, M.}, \bibinfo{author}{Mandel, O.},
  \bibinfo{author}{Esslinger, T.}, \bibinfo{author}{Hansch, T.~W.} \&
  \bibinfo{author}{Bloch, I.}
\newblock \bibinfo{title}{Quantum phase transition from a superfluid to a Mott
  insulator in a gas of ultracold atoms}.
\newblock \emph{\bibinfo{journal}{Nature}} \textbf{\bibinfo{volume}{415}},
  \bibinfo{pages}{39--44} (\bibinfo{year}{2002}).
\newblock \urlprefix\url{http://dx.doi.org/10.1038/415039a}.

\bibitem{0953-8984-20-12-123202}
\bibinfo{author}{Viefers, S.}
\newblock \bibinfo{title}{Quantum Hall physics in rotating Bose-Einstein
  condensates}.
\newblock \emph{\bibinfo{journal}{J. Phys.: Condens. Matter}}
  \textbf{\bibinfo{volume}{20}}, \bibinfo{pages}{123202}
   (\bibinfo{year}{2008}).
\newblock \urlprefix\url{http://stacks.iop.org/0953-8984/20/123202}.

\bibitem{PhysRevB.14.2239}
\bibinfo{author}{Hofstadter, D.~R.}
\newblock \bibinfo{title}{Energy levels and wave functions of Bloch electrons
  in rational and irrational magnetic fields}.
\newblock \emph{\bibinfo{journal}{Phys. Rev. B}} \textbf{\bibinfo{volume}{14}},
  \bibinfo{pages}{2239--2249} (\bibinfo{year}{1976}).

\bibitem{goldbaum:033629}
\bibinfo{author}{Goldbaum, D.~S.} \& \bibinfo{author}{Mueller, E.~J.}
\newblock \bibinfo{title}{Vortex lattices of bosons in deep rotating optical
  lattices}.
\newblock \emph{\bibinfo{journal}{Phys. Rev. A}}
  \textbf{\bibinfo{volume}{77}}, \bibinfo{pages}{033629}
  (\bibinfo{year}{2008}).
\newblock \urlprefix\url{http://link.aps.org/abstract/PRA/v77/e033629}.

\bibitem{umucalilar:055601}
\bibinfo{author}{Umucalilar, R.~O.} \& \bibinfo{author}{Oktel, M.~O.}
\newblock \bibinfo{title}{Phase boundary of the boson Mott insulator in a
  rotating optical lattice}.
\newblock \emph{\bibinfo{journal}{Phys. Rev. A}}
  \textbf{\bibinfo{volume}{76}}, \bibinfo{pages}{055601}
  (\bibinfo{year}{2007}).
\newblock \urlprefix\url{http://link.aps.org/abstract/PRA/v76/e055601}.

\bibitem{wu:043609}
\bibinfo{author}{Wu, C.}, \bibinfo{author}{dong Chen, H.},
  \bibinfo{author}{piang Hu, J.} \& \bibinfo{author}{Zhang, S.-C.}
\newblock \bibinfo{title}{Vortex configurations of bosons in an optical
  lattice}.
\newblock \emph{\bibinfo{journal}{Phys. Rev. A}}
   \textbf{\bibinfo{volume}{69}}, \bibinfo{pages}{043609}
  (\bibinfo{year}{2004}).
\newblock \urlprefix\url{http://link.aps.org/abstract/PRA/v69/e043609}.

\bibitem{tung:240402}
\bibinfo{author}{Tung, S.}, \bibinfo{author}{Schweikhard, V.} \&
  \bibinfo{author}{Cornell, E.~A.}
\newblock \bibinfo{title}{Observation of vortex pinning in Bose-Einstein
  condensates}.
\newblock \emph{\bibinfo{journal}{Phys. Rev. Lett.}}
  \textbf{\bibinfo{volume}{97}}, \bibinfo{pages}{240402}
  (\bibinfo{year}{2006}).
\newblock \urlprefix\url{http://link.aps.org/abstract/PRL/v97/e240402}.

\bibitem{mueller:041603}
\bibinfo{author}{Mueller, E.~J.}
\newblock \bibinfo{title}{Artificial electromagnetism for neutral atoms: Escher
  staircase and Laughlin liquids}.
\newblock \emph{\bibinfo{journal}{Phys. Rev. A}}
   \textbf{\bibinfo{volume}{70}}, \bibinfo{pages}{041603}
  (\bibinfo{year}{2004}).
\newblock \urlprefix\url{http://link.aps.org/abstract/PRA/v70/e041603}.

\bibitem{sorensen:086803}
\bibinfo{author}{Sorensen, A.~S.}, \bibinfo{author}{Demler, E.} \&
  \bibinfo{author}{Lukin, M.~D.}
\newblock \bibinfo{title}{Fractional quantum Hall states of atoms in optical
  lattices}.
\newblock \emph{\bibinfo{journal}{Phys. Rev. Lett.}}
  \textbf{\bibinfo{volume}{94}}, \bibinfo{pages}{086803}
  (\bibinfo{year}{2005}).
\newblock \urlprefix\url{http://link.aps.org/abstract/PRL/v94/e086803}.

\bibitem{JakschNJP2003}
\bibinfo{author}{Jaksch, D.} \& \bibinfo{author}{Zoller, P.}
\newblock \bibinfo{title}{Creation of effective magnetic fields in optical
  lattices: the Hofstadter butterfly for cold neutral atoms}.
\newblock \emph{\bibinfo{journal}{New J. Phys.}}
  \textbf{\bibinfo{volume}{5}}, \bibinfo{pages}{56} (\bibinfo{year}{2003}).
\newblock \urlprefix\url{http://stacks.iop.org/1367-2630/5/56}.

\bibitem{palmer-2008}
\bibinfo{author}{Palmer, R.~N.}, \bibinfo{author}{Klein, A.} \&
  \bibinfo{author}{Jaksch, D.}
\newblock \bibinfo{title}{Optical lattice quantum Hall effect}.
  \newblock \emph{\bibinfo{journal}{Phys. Rev. A}}
  \textbf{\bibinfo{volume}{78}}, \bibinfo{pages}{013609}
  (\bibinfo{year}{2008}).
\newblock
  \urlprefix\url{http://link.aps.org/abstract/PRA/v78/e013609}.

\bibitem{spielman:120402}
\bibinfo{author}{Spielman, I.~B.}, \bibinfo{author}{Phillips, W.~D.} \&
  \bibinfo{author}{Porto, J.~V.}
\newblock \bibinfo{title}{Condensate Fraction in a 2D Bose Gas Measured across the Mott-Insulator Transition}.
\newblock \emph{\bibinfo{journal}{Phys. Rev. Lett.}}
  \textbf{\bibinfo{volume}{100}}, \bibinfo{pages}{120402}
  (\bibinfo{year}{2008}).
\newblock \urlprefix\url{http://link.aps.org/abstract/PRL/v100/e120402}.

\bibitem{PhysRevLett.81.3108}
\bibinfo{author}{Jaksch, D.}, \bibinfo{author}{Bruder, C.},
  \bibinfo{author}{Cirac, J.~I.}, \bibinfo{author}{Gardiner, C.~W.} \&
  \bibinfo{author}{Zoller, P.}
\newblock \bibinfo{title}{Cold bosonic atoms in optical lattices}.
\newblock \emph{\bibinfo{journal}{Phys. Rev. Lett.}}
  \textbf{\bibinfo{volume}{81}}, \bibinfo{pages}{3108--3111}
  (\bibinfo{year}{1998}).

\bibitem{GretchenK.Campbell08042006}
\bibinfo{author}{Campbell, G.~K.} \emph{et~al.}
\newblock \bibinfo{title}{Imaging the Mott insulator shells by using atomic
  clock shifts}.
\newblock \emph{\bibinfo{journal}{Science}} \textbf{\bibinfo{volume}{313}},
  \bibinfo{pages}{649--652} (\bibinfo{year}{2006}).
\newblock
  \urlprefix\url{http://www.sciencemag.org/cgi/reprint/313/5787/649.pdf}.

\bibitem{folling:060403}
\bibinfo{author}{Folling, S.}, \bibinfo{author}{Widera, A.},
  \bibinfo{author}{Muller, T.}, \bibinfo{author}{Gerbier, F.} \&
  \bibinfo{author}{Bloch, I.}
\newblock \bibinfo{title}{Formation of spatial shell structure in the
  superfluid to Mott insulator transition}.
\newblock \emph{\bibinfo{journal}{Phys. Rev. Lett.}}
  \textbf{\bibinfo{volume}{97}}, \bibinfo{pages}{060403}
  (\bibinfo{year}{2006}).
\newblock \urlprefix\url{http://link.aps.org/abstract/PRL/v97/e060403}.

\bibitem{pu:190401}
\bibinfo{author}{Pu, H.}, \bibinfo{author}{Baksmaty, L.~O.},
  \bibinfo{author}{Yi, S.} \& \bibinfo{author}{Bigelow, N.~P.}
\newblock \bibinfo{title}{Structural phase transitions of vortex matter in an
  optical lattice}.
\newblock \emph{\bibinfo{journal}{Phys. Rev. Lett.}}
  \textbf{\bibinfo{volume}{94}}, \bibinfo{pages}{190401}
  (\bibinfo{year}{2005}).
\newblock \urlprefix\url{http://link.aps.org/abstract/PRL/v94/e190401}.

\bibitem{reijnders:060401}
\bibinfo{author}{Reijnders, J.~W.} \& \bibinfo{author}{Duine, R.~A.}
\newblock \bibinfo{title}{Pinning of vortices in a Bose-Einstein condensate by
  an optical lattice}.
\newblock \emph{\bibinfo{journal}{Phys. Rev. Lett.}}
  \textbf{\bibinfo{volume}{93}}, \bibinfo{pages}{060401}
  (\bibinfo{year}{2004}).
\newblock \urlprefix\url{http://link.aps.org/abstract/PRL/v93/e060401}.

\bibitem{PhysRevLett.86.4443}
\bibinfo{author}{Madison, K.~W.}, \bibinfo{author}{Chevy, F.},
  \bibinfo{author}{Bretin, V.} \& \bibinfo{author}{Dalibard, J.}
\newblock \bibinfo{title}{Stationary states of a rotating Bose-Einstein
  condensate: Routes to vortex nucleation}.
\newblock \emph{\bibinfo{journal}{Phys. Rev. Lett.}}
  \textbf{\bibinfo{volume}{86}}, \bibinfo{pages}{4443--4446}
  (\bibinfo{year}{2001}).

\bibitem{PhysRevB.20.1886}
\bibinfo{author}{Campbell, L.~J.} \& \bibinfo{author}{Ziff, R.~M.}
\newblock \bibinfo{title}{Vortex patterns and energies in a rotating
  superfluid}.
\newblock \emph{\bibinfo{journal}{Phys. Rev. B}} \textbf{\bibinfo{volume}{20}},
  \bibinfo{pages}{1886--1902} (\bibinfo{year}{1979}).

\bibitem{london}
\bibinfo{author}{London, F.}
\newblock \emph{\bibinfo{title}{Superfluids}} (\bibinfo{publisher}{Wiley},
  \bibinfo{address}{New York}, \bibinfo{year}{1990}).

\bibitem{1991LNP...394....1L}
\bibinfo{author}{{Leggett}, A.~J.}
\newblock \bibinfo{title}{{Topics in superfluidity and superconductivity}}.
\newblock In \bibinfo{editor}{{Hoch}, M.~J.~R.} \& \bibinfo{editor}{{Lemmer},
  R.~H.} (eds.) \emph{\bibinfo{booktitle}{Low Temperature Physics}}, vol.
  \bibinfo{volume}{394} of \emph{\bibinfo{series}{Lecture Notes in Physics,
  Berlin Springer Verlag}}, \bibinfo{pages}{1--92} (\bibinfo{year}{1991}).

\bibitem{rokhsar-1997}
\bibinfo{author}{Rokhsar, D.~S.}
\newblock \bibinfo{title}{Dilute Bose gas in a torus: vortices and persistent
  currents} (\bibinfo{year}{1997}).
\newblock
  \urlprefix\url{http://www.citebase.org/abstract?id=oai:arXiv.org:cond-mat/97%
09212}.

\bibitem{PhysRevA.57.R1505}
\bibinfo{author}{Mueller, E.~J.}, \bibinfo{author}{Goldbart, P.~M.} \&
  \bibinfo{author}{Lyanda-Geller, Y.}
\newblock \bibinfo{title}{Multiply connected Bose-Einstein-condensed
  alkali-metal gases: Current-carrying states and their decay}.
\newblock \emph{\bibinfo{journal}{Phys. Rev. A}} \textbf{\bibinfo{volume}{57}},
  \bibinfo{pages}{R1505--R1508} (\bibinfo{year}{1998}).

\bibitem{ryu:260401}
\bibinfo{author}{Ryu, C.} \emph{et~al.}
\newblock \bibinfo{title}{Observation of persistent flow of a Bose-Einstein
  condensate in a toroidal trap}.
\newblock \emph{\bibinfo{journal}{Phys. Rev. Lett.}}
  \textbf{\bibinfo{volume}{99}}, \bibinfo{pages}{260401}
  (\bibinfo{year}{2007}).
\newblock \urlprefix\url{http://link.aps.org/abstract/PRL/v99/e260401}.

\bibitem{PhysRevA.66.053606}
\bibinfo{author}{Kasamatsu, K.}, \bibinfo{author}{Tsubota, M.} \&
  \bibinfo{author}{Ueda, M.}
\newblock \bibinfo{title}{Giant hole and circular superflow in a fast rotating
  Bose-Einstein condensate}.
\newblock \emph{\bibinfo{journal}{Phys. Rev. A}} \textbf{\bibinfo{volume}{66}},
  \bibinfo{pages}{053606} (\bibinfo{year}{2002}).

\bibitem{jackson-2004-69}
\bibinfo{author}{Jackson, A.~D.}, \bibinfo{author}{Kavoulakis, G.~M.} \&
  \bibinfo{author}{Lundh, E.}
\newblock \bibinfo{title}{Phases of a rotating Bose-Einstein condensate with
  anharmonic confinement}.
\newblock \emph{\bibinfo{journal}{Phys. Rev. A}}
  \textbf{\bibinfo{volume}{69}}, \bibinfo{pages}{053619}
  (\bibinfo{year}{2004}).
\newblock
  \urlprefix\url{http://www.citebase.org/abstract?id=oai:arXiv.org:cond-mat/04%
01292}.

\bibitem{aftalion-2004-69}
\bibinfo{author}{Aftalion, A.} \& \bibinfo{author}{Danaila, I.}
\newblock \bibinfo{title}{Giant vortices in combined harmonic and quartic
  traps}.
\newblock \emph{\bibinfo{journal}{Phys. Rev. A}} \textbf{\bibinfo{volume}{69}},
  \bibinfo{pages}{033608} (\bibinfo{year}{2004}).
\newblock
  \urlprefix\url{http://www.citebase.org/abstract?id=oai:arXiv.org:cond-mat/03%
09668}.

\bibitem{jackson-2004-70}
\bibinfo{author}{Jackson, A.~D.} \& \bibinfo{author}{Kavoulakis, G.~M.}
\newblock \bibinfo{title}{Vortices in Bose-Einstein condensates with anharmonic
  confinement}.
\newblock \emph{\bibinfo{journal}{Phys. Rev. A}}
  \textbf{\bibinfo{volume}{70}}, \bibinfo{pages}{023601}
  (\bibinfo{year}{2004}).
\newblock
  \urlprefix\url{http://www.citebase.org/abstract?id=oai:arXiv.org:cond-mat/03%
11066}.

\bibitem{bargi-2006-73}
\bibinfo{author}{Bargi, S.}, \bibinfo{author}{Kavoulakis, G.~M.} \&
  \bibinfo{author}{Reimann, S.~M.}
\newblock \bibinfo{title}{Exact diagonalization results for an anharmonically
  trapped Bose-Einstein condensate}.
\newblock \emph{\bibinfo{journal}{Phys. Rev. A}}
  \textbf{\bibinfo{volume}{73}}, \bibinfo{pages}{033613}
  (\bibinfo{year}{2006}).
\newblock
  \urlprefix\url{http://www.citebase.org/abstract?id=oai:arXiv.org:cond-mat/05%
12503}.

\bibitem{scarola:210403}
\bibinfo{author}{Scarola, V.~W.} \& \bibinfo{author}{Das Sarma, S.}
\newblock \bibinfo{title}{Edge transport in 2d cold atom optical lattices}.
\newblock \emph{\bibinfo{journal}{Phys. Rev. Lett.}}
  \textbf{\bibinfo{volume}{98}}, \bibinfo{pages}{210403}
  (\bibinfo{year}{2007}).
\newblock \urlprefix\url{http://link.aps.org/abstract/PRL/v98/e210403}.

\bibitem{Fazio:2001ta}
\bibinfo{author}{Fazio, R.} \& \bibinfo{author}{van~der Zant, H.}
\newblock \bibinfo{title}{Quantum phase transitions and vortex dynamics in
  superconducting networks}.
\newblock \emph{\bibinfo{journal}{Physics Reports}}
  \textbf{\bibinfo{volume}{355}}, \bibinfo{pages}{235--334}
  (\bibinfo{year}{2001}).
\newblock
  \urlprefix\url{http://www.sciencedirect.com/science/article/B6TVP-44721TT-1/%
1/2b6c47028393ed5ecdd4f6ee7cbf5e83}.

\bibitem{Cataliotti:2001wc}
\bibinfo{author}{Cataliotti, F.~S.} \emph{et~al.}
\newblock \bibinfo{title}{Josephson junction arrays with Bose-Einstein
  condensates}.
\newblock \emph{\bibinfo{journal}{Science}} \textbf{\bibinfo{volume}{293}},
  \bibinfo{pages}{843--846} (\bibinfo{year}{2001}).
\newblock
  \urlprefix\url{http://www.sciencemag.org/cgi/content/abstract/293/5531/843}.

\bibitem{PhysRevLett.84.806}
\bibinfo{author}{Madison, K.~W.}, \bibinfo{author}{Chevy, F.},
  \bibinfo{author}{Wohlleben, W.} \& \bibinfo{author}{Dalibard, J.}
\newblock \bibinfo{title}{Vortex formation in a stirred Bose-Einstein
  condensate}.
\newblock \emph{\bibinfo{journal}{Phys. Rev. Lett.}}
  \textbf{\bibinfo{volume}{84}}, \bibinfo{pages}{806--809}
  (\bibinfo{year}{2000}).

\bibitem{ketterle-1999}
\bibinfo{author}{Ketterle, W.}, \bibinfo{author}{Durfee, D.~S.} \&
  \bibinfo{author}{Stamper-Kurn, D.~M.}
\newblock \bibinfo{title}{Making, probing and understanding Bose-Einstein
  condensates}. In
 \textit{Proceedings of the 1998 Fermi Summer School on BEC, Varenna, Italy} (IOS, Amsterdam, 1999).  
\newblock
  \urlprefix\url{http://www.citebase.org/abstract?id=oai:arXiv.org:cond-mat/99%
%04034}.

\bibitem{PhysRevB.40.546}
\bibinfo{author}{Fisher, M. P.~A.}, \bibinfo{author}{Weichman, P.~B.},
  \bibinfo{author}{Grinstein, G.} \& \bibinfo{author}{Fisher, D.~S.}
\newblock \bibinfo{title}{Boson localization and the superfluid-insulator
  transition}.
\newblock \emph{\bibinfo{journal}{Phys. Rev. B}} \textbf{\bibinfo{volume}{40}},
  \bibinfo{pages}{546--570} (\bibinfo{year}{1989}).

\bibitem{capogrosso-sansone:134302}
\bibinfo{author}{Capogrosso-Sansone, B.}, \bibinfo{author}{Prokof'ev, N.~V.} \&
  \bibinfo{author}{Svistunov, B.~V.}
\newblock \bibinfo{title}{Phase diagram and thermodynamics of the
  three-dimensional Bose-Hubbard model}.
\newblock \emph{\bibinfo{journal}{Phys. Rev. B }}
  \textbf{\bibinfo{volume}{75}}, \bibinfo{pages}{134302}
  (\bibinfo{year}{2007}).
\newblock \urlprefix\url{http://link.aps.org/abstract/PRB/v75/e134302}.

\bibitem{rey-2008}
\bibinfo{author}{Rey, A.~M.} \& \bibinfo{author}{Nunnenkamp, A.}
\newblock \bibinfo{title}{Macroscopic superposition states in rotating ring
  lattices} (\bibinfo{year}{2008}).
\newblock
  \urlprefix\url{http://www.citebase.org/abstract?id=oai:arXiv.org:0802.4309}.

\bibitem{PhysRevB.53.2691}
\bibinfo{author}{Freericks, J.~K.} \& \bibinfo{author}{Monien, H.}
\newblock \bibinfo{title}{Strong-coupling expansions for the pure and
  disordered Bose-Hubbard model}.
\newblock \emph{\bibinfo{journal}{Phys. Rev. B}} \textbf{\bibinfo{volume}{53}},
  \bibinfo{pages}{2691--2700} (\bibinfo{year}{1996}).

\bibitem{PhysRevLett.87.060403}
\bibinfo{author}{Ho, T.-L.}
\newblock \bibinfo{title}{Bose-Einstein condensates with large number of
  vortices}.
\newblock \emph{\bibinfo{journal}{Phys. Rev. Lett.}}
  \textbf{\bibinfo{volume}{87}}, \bibinfo{pages}{060403}
  (\bibinfo{year}{2001}).

\end{thebibliography}

\end{document}